\documentclass[preprint,12pt,authoryear]{elsarticle}

\usepackage{amssymb}
\usepackage{amsmath}

\usepackage{longtable}
\usepackage{tabularx}
\usepackage{booktabs}
\usepackage{url}

\journal{Preprint}

\begin{document}

\begin{frontmatter}



\title{Spatial Activity Analysis of Neuromuscular Junctions in Three-Dimensional Histology-based Muscle Reconstructions} 


\author[ECE]{Alessandro Ascani Orsini} 
\author[BME]{Manan Bhatt}
\author[BME]{Pierce L. Perkins}
\author[BME]{Siyu Wang}
\author[BME]{Kiara N. Quinn}
\author[BME]{Kenzi Griffith}
\author[BME]{Fausto Kang}
\author[ECE,BME]{Nitish V. Thakor}

\affiliation[ECE]{organization={Department of Electrical and Computer Engineering, Johns Hopkins University},
            city={Baltimore},
            state={MD},
            country={USA}}
\affiliation[BME]{organization={Department of Biomedical Engineering, Johns Hopkins University},
            city={Baltimore},
            state={MD},
            country={USA}}

\begin{abstract}
Histology has long been a foundational technique for studying anatomical structures through tissue slicing. Advances in computational methods now enable three-dimensional (3D) reconstruction of organs from histology images, enhancing the analysis of structural and functional features. Here, we present a novel multimodal computational method to reconstruct rodent muscles in 3D using classical image processing and data analysis techniques, analyze their structural features and correlate them to previously recorded electrophysiological data. The algorithm analyzes spatial distribution patterns of features identified through histological staining, normalizing them across multiple samples. Further, the algorithm successfully correlates spatial patterns with high-density epimysial ElectroMyoGraphy (hdEMG) recordings, providing a multimodal perspective on neuromuscular dynamics, linking spatial and electrophysiological information. The code was validated by looking at the distribution of NeuroMuscular Junctions (NMJs) in naïve soleus muscles and compared the distributions and patterns observed with ones observed in previous literature. Our results showed consistency with the expected results, validating our method for features and pattern recognition. The multimodal aspect was shown in a naïve soleus muscle, where a strong correlation was found between motor unit locations derived via hdEMG, and NMJ locations obtained from histology, highlighting their spatial relationship. This multimodal analysis tool integrates 3D structural data with electrophysiological activity, opening new avenues in muscle diagnostics, regenerative medicine, and personalized therapies where spatial insights could one day predict electrophysiological behavior — or vice versa.
\end{abstract}

\begin{graphicalabstract}
\includegraphics[width=1.0\linewidth]{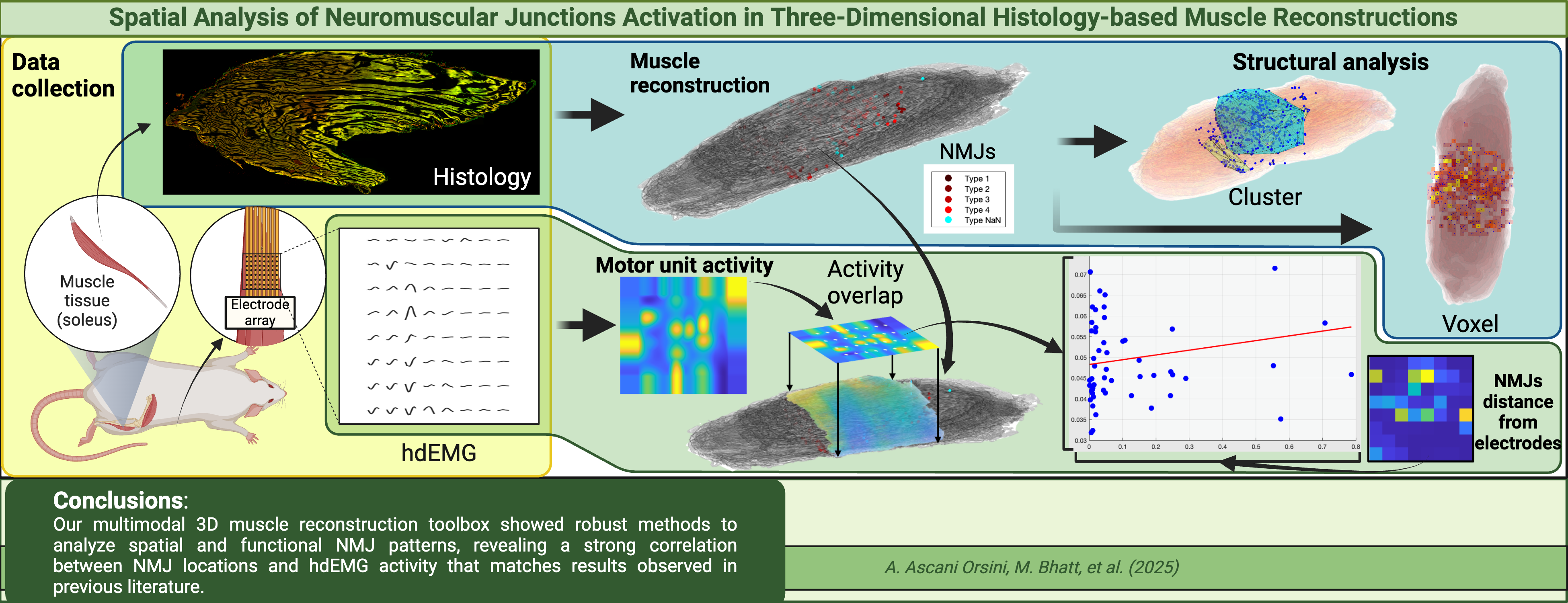}
\end{graphicalabstract}

\begin{highlights}
\item Developed a novel computational pipeline to reconstruct rodent muscle tissues in 3D from histology images, enabling detailed visualization and analysis of structural patterns, such as NMJ distribution.
\item The structural analysis of NMJ patterns in naïve soleus muscles showed consistency with traditional patterns observed in the literature.
\item The algorithm demonstrated its ability to correlate structural feature patterns with hdEMG data, bridging structural and electrophysiological information to provide a detailed analysis of how changes in one can influence the other, supporting histology-based research focused on muscle.
\end{highlights}

\end{frontmatter}



\section{Introduction}
\label{intro}
Histology has been around since the early 19th century as a tool to understand the morphology and anatomy of organs and cells within them (\cite{mazzarini2021evolution}). Often, this technique involves the use of labels to highlight structures of interest to analyze and observe over different animals. Traditionally, this technique produces two-dimensional slides which can obscure or distort the complex three-dimensional relationships of the data in the tissues (\cite{prajapati2016acquisition,ford2023new}). Further, variations in animal samples generate diverse data sets, making it difficult to find common patterns in the samples. Histology is primarily a morphological analysis technique, offering limited insight into the functional activity of the structures. As a result, it provides only a partial perspective, making it challenging to integrate findings with other recordings and observations. This limitation can lead to incomplete conclusions or overlooked observations due to the inability to visualize both dimensions jointly (\cite{ruusuvuori2022spatial}). In muscle studies, it is often important to investigate not only the structural changes observed during experiments but also the electrophysiological activity of the tissue. Establishing a link between these two aspects could prove highly beneficial, for instance by identifying which NMJs are functional and which are not, mapping patterns of activation in histology structures, and understanding how these identified patterns can change based on the experiments performed.
\\\\
3D histology systems have emerged to provide dimensionality to histology methods, providing important insights in understanding tissue morphology, connectivity and cellular composition(\cite{van20223d, carragher2018concerns, ruusuvuori2022spatial}). More recent attempts have been trying to merge these histology reconstructions with other types of three dimensional data such as from CT or MRI \cite{morawski2018developing,straight2024histological}, this has opened venues to start combining not only structural data but also functional data in one individual model. However, despite these advancements, challenges remain in fully capturing the interplay between structure and function, specifically with data from low-dimensional systems (such as two- and one-dimensional systems) \cite{kwak2019deephealth,blini2024pupillary}. In particular, integrating dynamic electrophysiological measurements with static structural reconstructions is not straightforward, often leading to fragmented insights and complex modelling \cite{fernandez2016multiscale, bucelli2023mathematical}. This is especially significant in muscle studies, where both anatomical and functional information are essential. 
\\\\
Thus, there is a clear need for a method that can address these limitations, leverage the gathered histology of muscle tissue to reconstruct the organs in three dimensions and identify specific elements in it. The paper at hand addresses these limitations by introducing a multimodal computational pipeline that integrates 3D structural reconstructions and analysis with electrophysiological data, providing a more holistic and functionally relevant perspective on tissue organization. Such a solution uses various staining techniques to emphasize features of interest and facilitate pattern recognition within data distributions. Furthermore, this algorithm integrates the histological data with the muscle hdEMG activity recorded, adding an additional layer of analysis between these characteristics and muscle activation patterns. This integration would create a comprehensive model that not only depicts the muscle's structure but also links it to its function, permitting a better understanding of how one influences the other across multiple experimental conditions. Such technology could provide deeper insights into the correlations between different structures within an organ and the effects of one physiological signal on others.
\\\\
This method will be used to analyze and compare the naïve soleus muscles of rodents with the traditionally observed results from the literature (\cite{carre2022distribution,cardasis1981ultrastructural,luff1998age}).
The program will focus on observing the location of NMJs and their distribution across multiple samples and analyze how their location is related to the activity measured in vivo.  
\\\\
This paper will focus on detailing and demonstrating the methodology used to analyze feature patterns and distributions in histology slides of naïve rodent soleus muscles. It will include generating their 3D reconstructions, identifying trends in a feature of interest (NMJs), by looking at its distribution and clustering, and correlating these patterns with in vivo hdEMG recordings. We will then discuss the observations of the application of these methods and their limitations, concluding with potential directions for future research in this promising field.
\section{Materials and Methods}
In our study, we developed an integrated computational framework that combines histology, three-dimensional reconstruction, and electrophysiological data analysis to investigate NMJ activation. Muscle tissue sections were prepared using standard fixation and staining protocols to ensure clear visualization of both muscle fibers and NMJs. High-resolution histological images underwent pre-processing, including noise reduction and alignment. NMJ identification was performed manually and overlaid onto the corresponding histology slides. After adjustment, the serial images were compiled into volumetric models using a dedicated 3D reconstruction algorithm, preserving the spatial context of NMJs within the muscle tissue.

We then conducted a NMJ distribution analysis using both a voxel- and density-based clustering method to quantify key parameters such as NMJ density, clustering, and spatial dispersion. Additionally, we correlated NMJ spatial distribution with electrophysiological activity recorded in vivo from the muscle during mechanical stimulation. Statistical correlation techniques were applied to validate the relationship between NMJ density and distribution metrics with EMG activity, linking the structural organization of NMJs to their functional electrical output.
\label{materials}
\subsection{Histology and feature aquisition}
\label{histo}
To test our algorithm we used soleus muscles from Lewis rats. The samples were fixed right after harvesting with 4\% paraformaldehyde for histology analysis. After a 24-hour fixation period, the tissue samples were cryoprotected using 15\%, followed by 30\% sucrose solutions (48 hours each). The tissue was then frozen in optimal cutting temperature compound, sectioned longitudinally (20 µm thickness). Staining of the rodent soleus muscle was performed using beta-III tubulin (1:800 in PBS and goat serum, Invitrogen) to label axons and alpha-bungarotoxin (1:800 in PBS and goat serum, Invitrogen) to label NMJs to analyze the location of the NMJs within the muscle. Every 8th section was imaged and analyzed using Leica DMi8 (160 µm). The coordinates of the junctions were manually found by two independent rating experts using FIJI software and were classified into four different categories according to their level of reinnervation as described in \cite{vannucci2019normal}. All the codes that were used in this work are available on the Github Repository provided in the Data Availability section of this paper. All variables and their descriptions are reported in Table \ref{tab:physiological-variables} subdivided by code they are used in.

\subsection{Reconstruction algorithm}
\label{reconstruction}
\begin{figure}[t]
\centering
\includegraphics[width=1.0\linewidth]{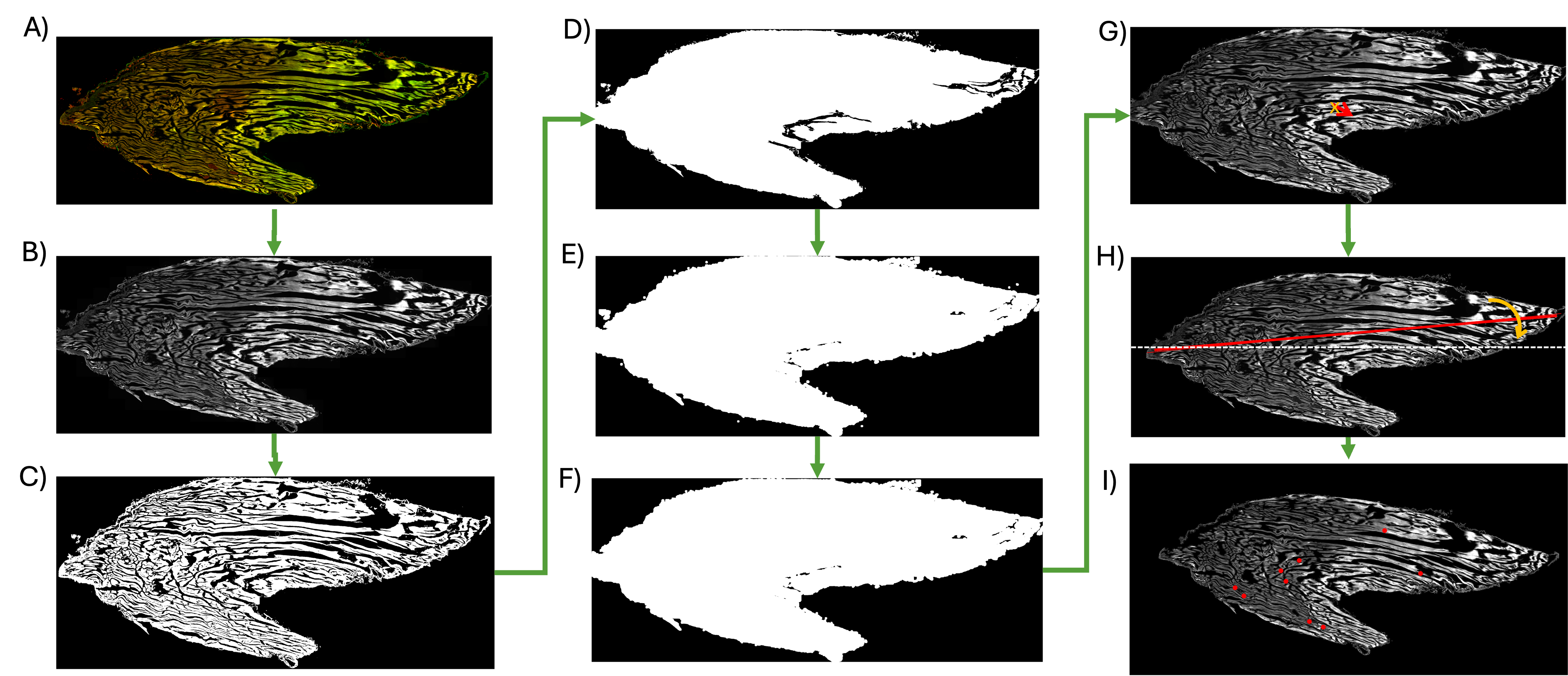}
\caption{\textbf{Image processing pipeline from the histology slide.} A) Original stained histology slide of rat naïve soleus. B) Greyscale conversion of the slide. C) Mask obtained via Otsu thresholding of the image. D) Flood fill hole covering of the mask E) Dilatation of the mask via oval structuring element F) Selection of the largest component area in the mask G) Original image filtered using the mask, and centroid shift to the center. H) Rotation of the image about the x axis (white dashed line) and the centre based on mean angle (orange dashed line) of the feature lines (red lines). I) Plotting of the updated NMJ coordinates on the slide.}\label{fig:flow}
\end{figure}

To reconstruct the organ, histology images are processed using MATLAB (algorithm \texttt{analysis.m}). The program imports the images along with NMJ coordinates and classifications (Figure \ref{fig:flow}A). The slides are then converted to greyscale and rescaled to enhance code efficiency and performance (Figure \ref{fig:flow}B).

The next step involves removing the background and isolating the slide’s shape. This is achieved by applying a mask based on a threshold determined using the Otsu method (\cite{otsu1979}), which separates regions by maximizing inter-class variance (Figure \ref{fig:flow}C). To remove potential gaps left within the sample, a flood-fill algorithm was implemented to identify and close holes within the mask (Figure \ref{fig:flow}D). In some cases, the Otsu method separated the tissue into distinct regions, thus ensuring all parts were recognized as part of the same sample, an oval structuring element was applied with a radius of 10 pixels, slightly dilating the shapes to connect them (Figure \ref{fig:flow}E).

Once all shapes in the mask are connected, the program segments the mask into regions using an 8-connectivity connected-components labeling algorithm (\cite{Maurer2003}). The regions are then analyzed by label, and the largest component by area—representing the tissue—is identified and isolated (Figure \ref{fig:flow}F). The mask’s centroid is subsequently extracted and used to reposition the entire slide to the center of the frame (Figure \ref{fig:flow}G). To avoid cropping during this shift, the image border is increased based on the centroid’s movement.

Histology slides often lack proper orientation, with alignment varying between images. To correct this, the algorithm prompts the user to manually identify at least one orientation line usually based on structural elements in the tissue, such as muscle fiber orientation or scar tissue. The program then calculates the average angle between the identified orientation lines highlighted by the user, using it to rotate both the image and the mask around their center counterclockwise and align them with the horizontal axis (Figure \ref{fig:flow}H). Notably, the angle is determined as the one between the vector defined by the first and second points input by the user.

Once all histology slides are processed, the program stacks them along the Z-axis, maintaining a spacing proportional to the slice thickness (20 $\mu$m) multiplied by the missing number sections (8) and applying an alpha value to introduce transparency. The convex hull of the stacked slide contours is then calculated and overlaid to approximate the muscle’s profile and 3D shape, providing a clearer representation of its spatial configuration.

\subsection{Extracting muscle features}
\label{Muscle features}
To track changes in muscle tissue across multiple experiments, extracting key metrics such as volume and length becomes essential, specifically to obtain meaningful comparisons across animals of the same groups. These metrics can be derived from the 3D reconstruction and allow normalization of tissue samples between animals within the same group and provide insight into aspects such as the atrophy of the tissue.

\subsection{Overlaying NMJ in the 3D space}
\label{NMJs}
The next step after reconstructing the muscle is to map the type and location of the NMJs. The coordinates are processed using the same transformer applied earlier to align, scale, and rotate the histology slide to the origin. These coordinates are then color-coded based on features such as fiber typology or junction maturation level. Finally, the coordinates are overlaid on the muscle 3D model, illustrating their spatial distribution (Figure \ref{fig:flow}I).

\subsection{Muscle volume estimation}
\label{volume}
\begin{figure}[t]
\centering
\includegraphics[width=0.8\linewidth]{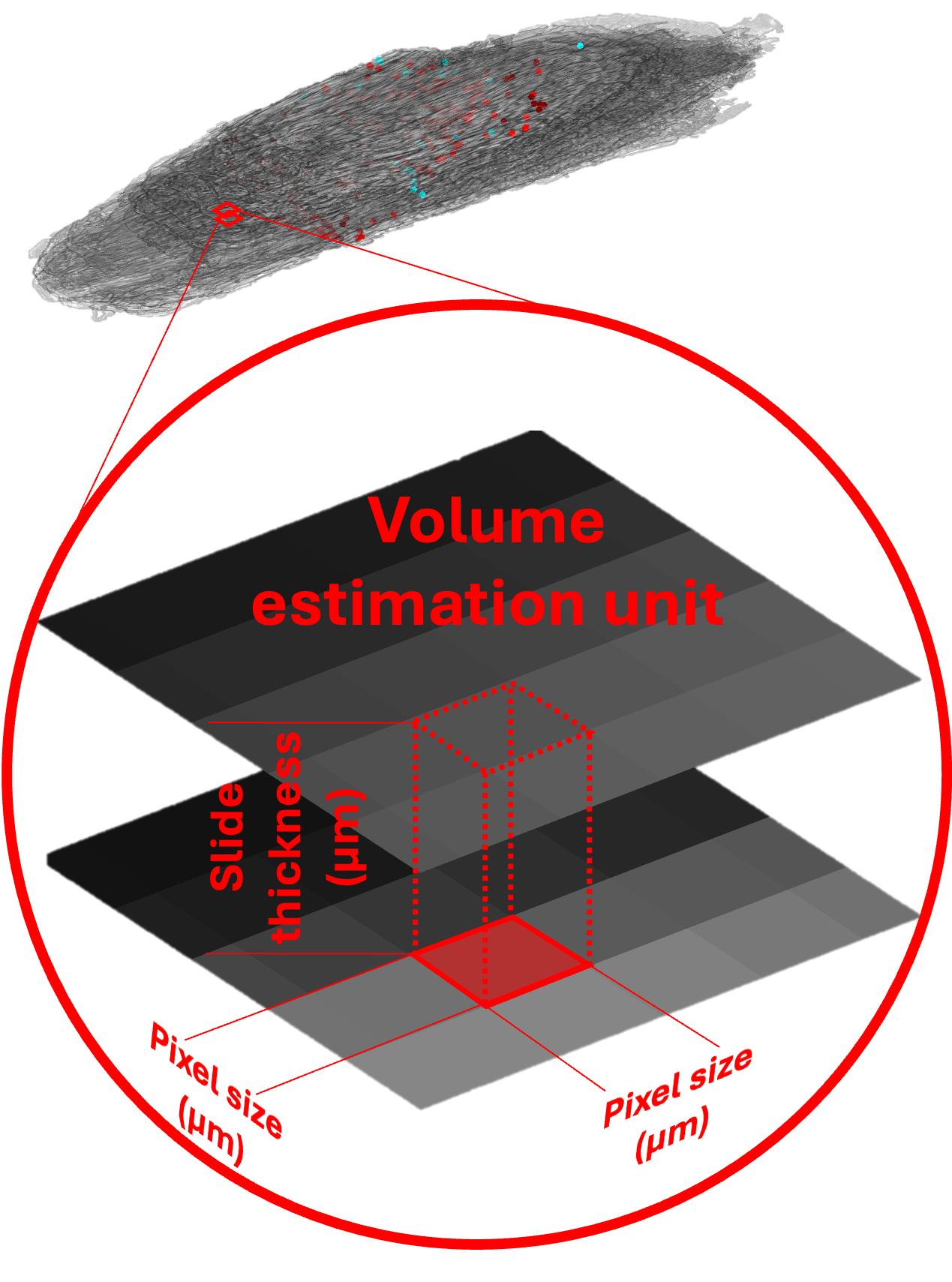}
\caption{\textbf{Volume estimation method} Muscle volume is estimated using a reconstruction approach based on histological slide data. Each individual unit of volume is modeled as a cuboid, where the side lengths are determined by the in-plane pixel dimensions in micrometers ($\mu$m), and the height corresponds to the thickness of each histological slide. By summing these unit volumes across all pixels that constitute the muscle reconstruction, a total volume estimate is obtained. This method provides a voxel-based approximation of muscle tissue volume, allowing for quantitative comparisons across different samples.}\label{fig:volEst}
\end{figure}

To reliably estimate muscle volume, the volume of each pixel in a slide is calculated and multiplied by the total number of pixels that form the shape of the muscle. The pixel volume was approximated as a rectangular cuboid, with the height equal to the slice thickness and the side lengths obtained from the pixel dimensions indicated by the microscope (Leica DMi8). The structure of a single unit of volume is showed in Figure \ref{fig:volEst}. Once the unit volume is estimated it is multiplied by the number of pixels in the figure. This process is then repeated for all Z-stacked layers, and the results are summed to provide an approximate total muscle volume. 

\subsection{Estimation of length of the muscle}
\label{length}
The longitudinal length of the muscle is estimated from histology slides using a rotational axis method. This approach involves rotating a line passing through the center of the 2D image by 180 degrees in 0.1-degree increments. At each step, the algorithm calculates the longest distance between the line’s intersections with the muscle contour on the histology slide. The maximum distance across all slides is then designated as the muscle’s longitudinal length.

\subsection{Normalization of the data}
\label{norm}
\begin{figure}[t]
\centering
\includegraphics[width=1.0\linewidth]{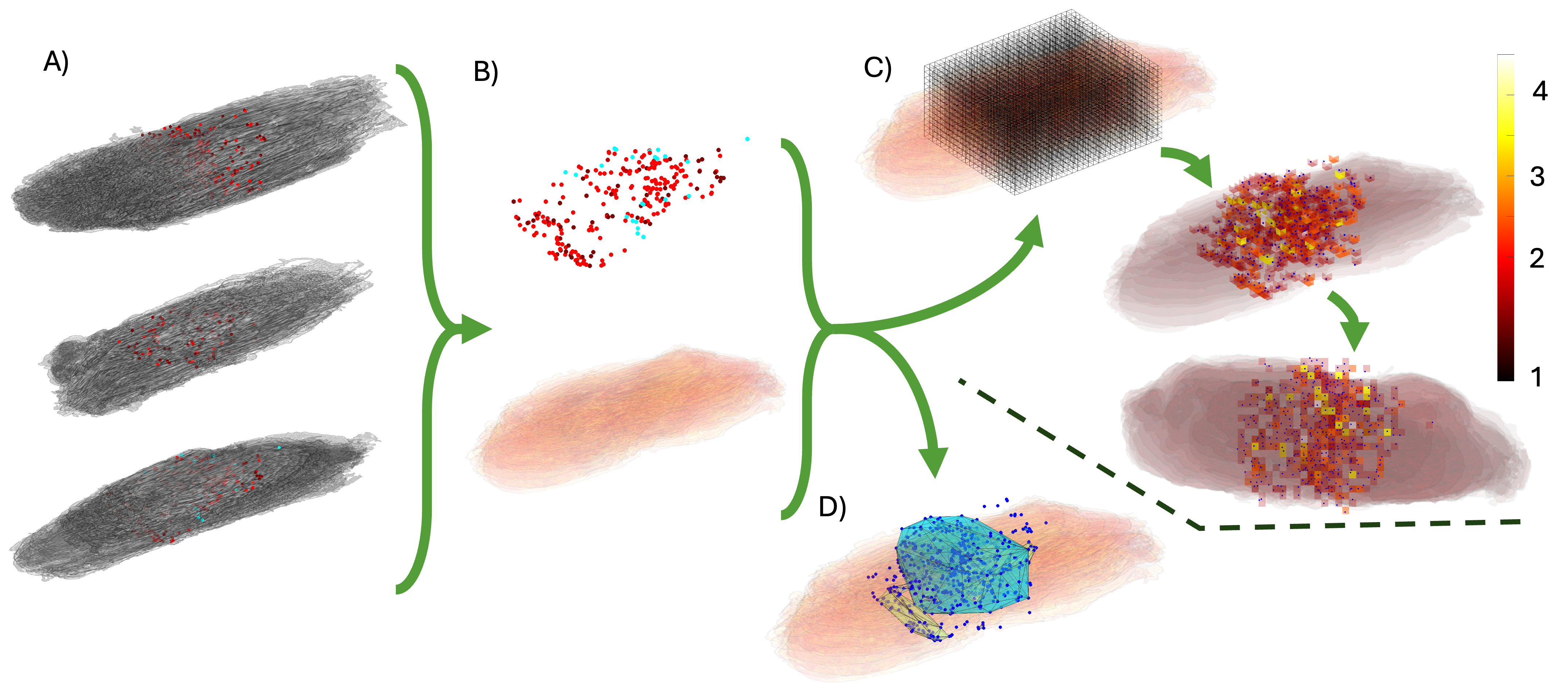}
\caption{\textbf{NMJ distribution analysis techniques.} A) After Z stacking and plotting the NMJs coordinates of different muscles, we can perform different averaging methods. The following shown are Naïve soleus muscles. B) From the raw muscle reconstruction we can extract the oriented NMJ coordinates and the muscle models. A model muscle can then be selected to give a sense of spatiality and anatomy and the coordinates extracted from all the muscles can be normalized based on the muscle's dimensions. C) Averaging method using voxel based analysis (Section \ref{voxel}). The muscle is divided into a grid with a given voxel size. The color of the voxel indicates the number of NMJ contained within and the blue dot is the average coordinate of the junctions present within. D) Cluster finding method based on density scan}\label{fig:flow2}
\end{figure}

To analyze structural trends in NMJ distribution across muscles (algorithm \texttt{Avg\_NMJ\_contour.m}) it is necessary to normalize the data. To compare the same muscle across multiple animals, we employed Procrustes Analysis (\cite{gower1975generalized}) because it effectively isolates true shape differences by removing variations due to translation, rotation, and scaling. In contrast, simpler methods such as direct point-to-point comparisons or z-score normalization do not adjust for geometric misalignment. Point-to-point methods compare raw coordinates without aligning the structures, meaning that differences in orientation or size can obscure genuine morphological variations (\cite{liu2023analysis}). Similarly, while z-score normalization standardizes individual feature values, it fails to consider the overall spatial configuration of the muscle, which can lead to an exaggerated sensitivity to local variations (\cite{liu2023analysis}). Thus, Procrustes Analysis offers a more robust framework for accurately capturing and comparing the spatial patterns and morphological features of the muscle. Subsequenly, a representative muscle is chosen as the model to identify and visualize the location of NMJs within its structure. The program then loads the 3D reconstruction of all the muscles to be compared (Figure \ref{fig:flow2}A). For each, it extracts the NMJs coordinates and the muscle boundaries in Cartesian coordinates (Figure \ref{fig:flow2}B).

The program then scales uniformly the coordinates of each muscle based on the ratio between each dimension of the original muscle and the model muscle. This scaling ensures that the coordinates remain within the boundaries of the model muscle across all axes. Additionally, the scaled positions are proportionally aligned with their original locations, maintaining spatial fidelity in the reconstructed model.

\subsection{Voxel averaging method}
\label{voxel}
To visualize the distribution and density of NMJs, the program segments the region containing NMJs into voxels of a specified size (20 $\mu$m in this case). It iterates through each coordinate, determines the corresponding voxel, and counts the number of NMJs within it. The program then averages the coordinates of NMJs within each voxel and reconstructs the Z-stack of the model muscle. This reconstruction is overlaid with a three-dimensional grid representing the voxel locations, where each voxel is color-coded based on the number of NMJs it contains, with the average coordinate plotted within. Figure \ref{fig:flow2}C illustrates an example of 3D heatmap of the NMJ. The angle of orientation of the NMJs along the saggital plane of the muscle was determined using Principal component analysis, by obtaining the eigenvector of all the averaged NMJ locations.

\subsection{Density Based Clustering}
\label{dbscan}
To identify clustering trends within the muscle, we use a density-based approach with the 3D data based on the method described in \cite{ester1996}. The code runs by examining each NMJ within a radius $\epsilon$ and considers points as core points if they have more than $n$ neighbors within the radius. Core points within $\epsilon$ of each other connect to form clusters, and non-core points are added to these clusters if they also fall within the radius. The sample results of this approach on the coordinates are shown in Figure \ref{fig:flow2}D.

\subsection{Overlaying activity heatmap with muscle histology}
\label{overlay}
\begin{figure}[t]
\centering
\includegraphics[width=1.0\linewidth]{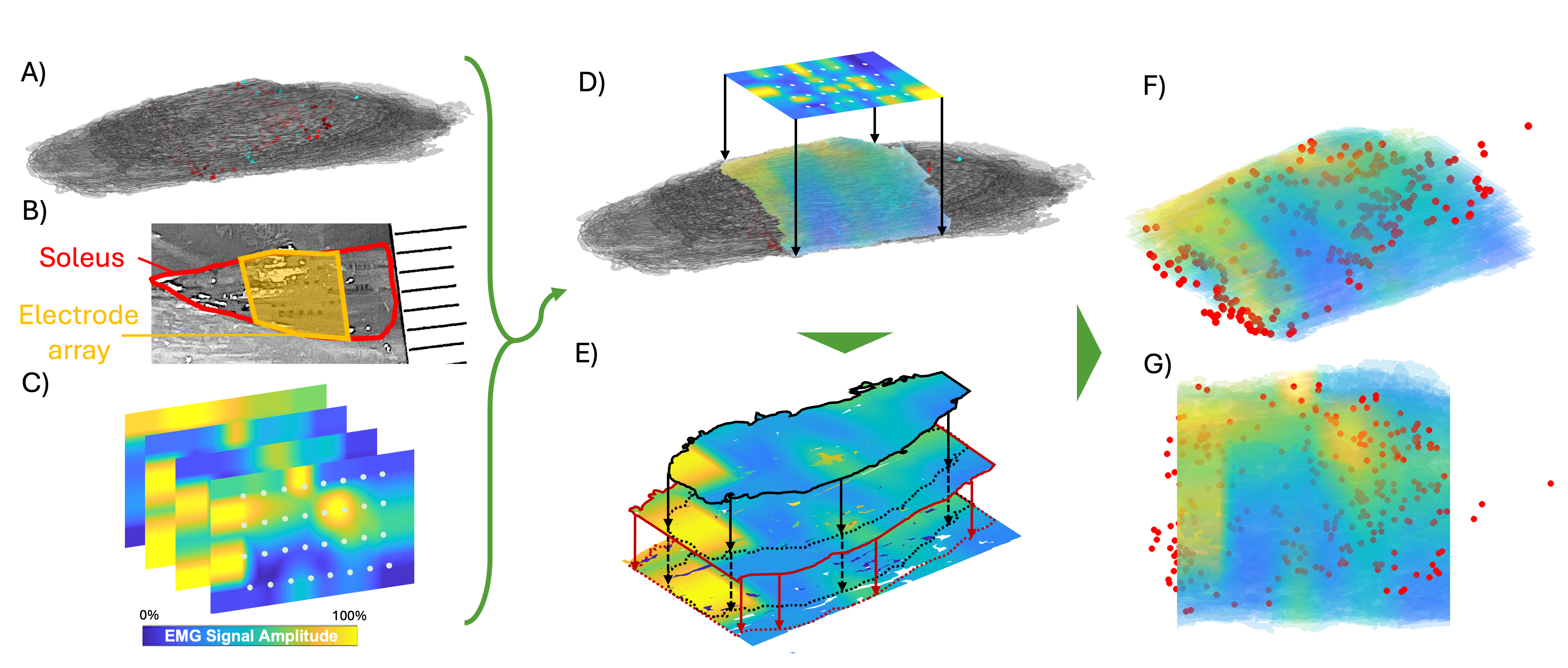}
\caption{\textbf{Schematic showing how the EMG activity from the electrodes was overlapped with the histology reconstruction of the muscle.} A) The reconstructed muscle and its data are one of the inputs of the process. B) Photo of the muscle and the electrodes placement during the recording. This image is important to obtain the placement location of the electrode, its covered area on the muscle and the orientation of the array. Scale on the side is in mm. C) Interpolated EMG heatmap recorded over time from the electrode array. In this case, the data recorded correspondeed with a mechanical stimulation test used to activate as many motor units as possible. D) Based on the location and the respectful electrode area, the EMG recordings are scaled, aligned and overlayed with the muscle reconstruction. The goal, is to 'wrap' the heatmap around the muscle while also considering the signal transmission through the tissue. E) Sample slides of the tissue showing how the 'wrapping' around the muscle is simulated. The original tissue is used as mask applied to the heatmap. Further, the already masked layers above the current one, are subtracted, reducing the values within the heatmap by a constant $\alpha$ meant to represent the electric loss of the signal within the tissue. F) Perspective view of the overlayed EMG activity stacked with the location of the NMJs within the muscle. G) Top view of the overlayed EMG activity.}\label{fig:OverlapExplained}
\end{figure}

To better understand muscle function, rehabilitation after injury, or, in our case, changes in structural and functional distribution, it is important to link the spatial information obtained from histology (Figure~\ref{fig:OverlapExplained}A) with the electrophysiology data collected in vivo using hdEMG. This can provide additional insights into how the physical feature  contributes to the muscle’s overall function.

To do so we used the myomatrix (\cite{myomatrix}), a 32-channel electrode array, to record EMG activity from the soleus with high spatial and temporal resolution. A picture of the muscle and the electrode placement is taken for later estimation of the area of the muscle recorded by the electrodes 
as shown in figure \ref{fig:OverlapExplained}B. Mechanical stimulation is applied to the sciatic nerve using forceps, activating all motor units in the muscle to visualize their activation locations. Through interpolation and kilosort peak detection, a heatmap is generated for each 200 ms recording interval (Figure \ref{fig:OverlapExplained}C).

The activity heatmap is scaled and aligned with the histological NMJ reconstruction by mapping it to the electrode array’s starting and ending positions, determined from the reference image (Figure\ref{fig:OverlapExplained}B). The activity map is then layered onto the histological reconstruction in a blanket-like manner (Figure \ref{fig:OverlapExplained}D), where the highest activity corresponds to the most superficial parts of the muscle.

To overlay the activity heatmap with the histological NMJ reconstruction, the heatmap is scaled and aligned based on the starting and ending positions of the electrode array, as determined from the photograph (Figure \ref{fig:OverlapExplained}B). The activity map is then layered onto the histological reconstruction like a blanket (Figure \ref{fig:OverlapExplained}D), with the highest activity corresponding to the most deeper parts of the muscle. Understanding the spatial distribution of the signal within the muscle is crucial, so we use an exponential decay model to simulate signal attenuation in lossy tissue (\cite{alabaster2004microwave,ulaby2015fundamentals}):

\[V(d) = V_0 e^{-\beta d} = V_0 (1-\alpha)^d, \]
\[\beta = \sqrt{\pi\mu f(\sigma+j2\pi f\epsilon)}, \]
\[\alpha = 1-e^{-\beta}, \]

Where the real part of $\alpha$ is a damping constant used in the program to exponentially decrease signal intensity each layer. $\alpha$ is kept negative since the reconstruction happens from the epimysium of the muscle to its deeper regions.  With $\sigma$ = 0.3 S/m, $\epsilon = 10^7$ (\cite{miklavvcivc2006electric}), $\mu$ was approximated to $\mu_0$ and a frequency of maximum 200 Hz gave an $\alpha = -0.015$ which was used for the simulation. This was applied by using the muscle histology as mask for the activity heatmap at its original intensity as represented in figure \ref{fig:OverlapExplained}E. 
Once the intensity of all the layers was calculated, the NMJs were plot in the EMG reconstruction to observe the overlay with heatmap activity (Figure \ref{fig:OverlapExplained}F-G). (\cite{reaz2006techniques})

\subsection{NMJ distribution and hdEMG reconstruction}
\label{hdEMG}
\begin{figure}[t]
\centering
\includegraphics[width=1.0\linewidth]{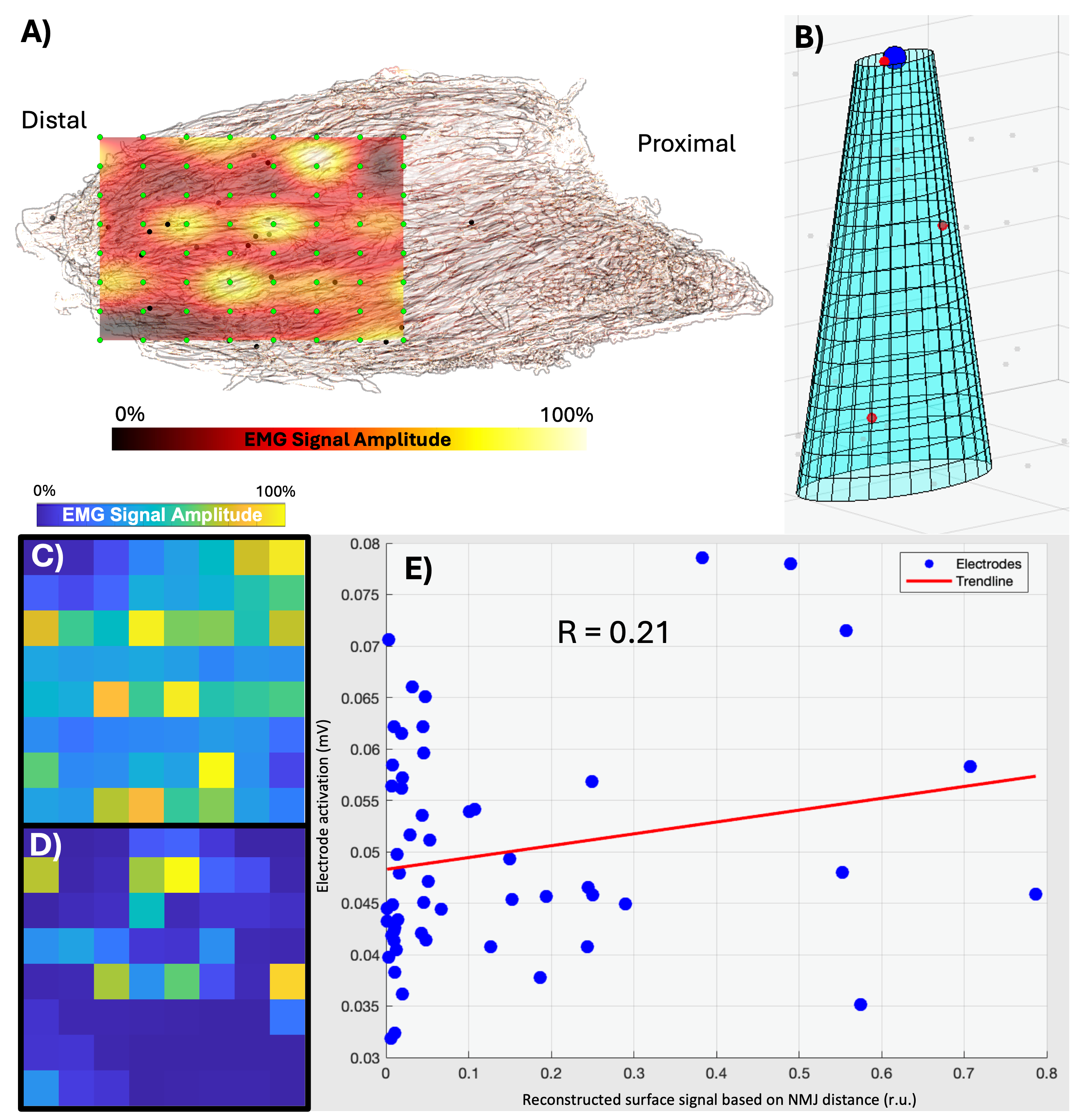}
\caption{\textbf{Correlation analysis of epimysial activity and NMJ distance.} A)EMG activity heatmap overlapped on oriented naïve soleus muscle. Green dots indicate electrode location and black dots NMJ location B) Inverse cone recording region under an electrode C) Overlayed EMG signal from the electrode positions D) Reconstructed signal normalized and modeled based on the relative distance of each NMJ to the corresponding electrode in the conical structure E) Linear regression analysis comparing the recorded electrode activity to the reconstructed signal derived from NMJ distances. The regression line for each electrode quantifies the relationship, with R representing the Pearson correlation coefficient, indicating the strength and direction of the association.}\label{fig:overlapEMG}
\end{figure}

To more robustly link NMJs anatomical features with recorded muscle activity—and to show the method’s potential for cross-validating histological and electrophysiological changes we correlated NMJ locations to areas of high activity on the epimysium. This was done with two main assumptions: 1) the depth of the NMJs affects the signal on the epimysium 2) Each electrode records from a cone shaped area right under it and 3) All the NMJs included in the area of the cone of the electrode contribute to the epymisial signal. To test this, we first used the alignment from our previous method to define the electrode placement area (Figure \ref{fig:overlapEMG}A). The activity of the individual electrodes was derived then from the uninterpolated heatmap. The cone of each electrode had its top radius as half the inter-electrode distance and its base radius as equal to the full electrode distance (Figure \ref{fig:overlapEMG}B). For the activity of the NMJs, we applied a signal decay model incorporating a reverse damping constant ($\alpha = 0.015$), reflecting the fact that the reconstruction originates at the NMJs rather than the epimysium. By summing the decayed contributions at the muscle surface, we simulated the signal that would be recorded by the electrode. Finally, we used Pearson correlation and regression analyses to assess the relationship between NMJ depth and the recorded signal strength (Figure \ref{fig:overlapEMG}C-E).

\section{Results}
\label{results}
\begin{figure}[t]
\centering
\includegraphics[width=0.8\linewidth]{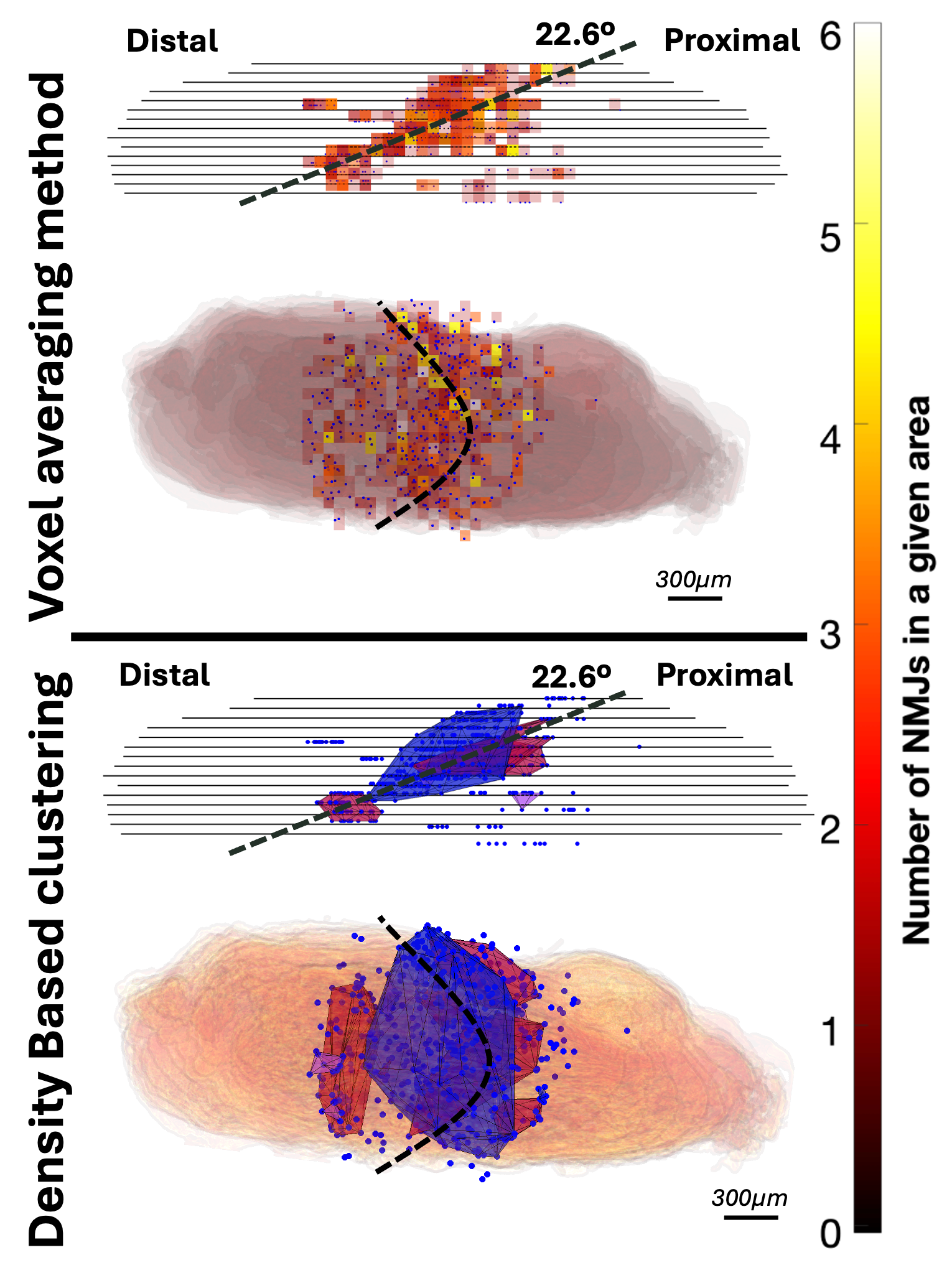}
\caption{\textbf{Structural differences in spatial features across analysis methods} This figure illustrates the analysis results from the voxel averaging method (Section \ref{voxel}) and density-based clustering (Section \ref{dbscan}), comparing a naïve soleus to a vRPNI. Both methods reveal significant structural differences between the two surgical models and consistent patterns across animals within the same group (n=3). Both approaches used in the program offer valuable options to analyze, quantify and compare structural features across different experimental conditions in muscle tissue.}\label{fig:results}
\end{figure}

The structural analysis performed by our method reveals visible similarities between the patterns of NMJs in naïve rodent soleus muscles and the ones observed in the literature, with the observable ±20 degrees orientation and the "zigzag" pattern described in \cite{carre2022distribution,cardasis1981ultrastructural,luff1998age}, as shown in Figure \ref{fig:results}. In addition to the structural analysis, our program demonstrated consistency in mapping NMJ spatial distributions with corresponding electrophysiological activity. 

The voxel averaging method (Section \ref{voxel}) across 3 muscles, highlighted regions of high NMJ density concentrated centrally within the muscle, with patterns more focused in the middle region of the muscle (Figure \ref{fig:results} Top). The density-based clustering algorithm (Section \ref{dbscan}) further refined this visualization, pinpointing dominant clusters of NMJs along the central axis, indicating features of nerve branching (Figure \ref{fig:results} Bottom). The overlaying of hdEMG data sets revealed localized areas of high activity coinciding with the spatial regions containing dense NMJ clusters (Figure \ref{fig:overlapEMG}A). This was supported by the idea that NMJ location influences the observed electrical signals. The Pearson correlation analysis between the activity and the NMJ location using an lossy tissue decay model yielded an R-value of 0.21 (Figure \ref{fig:overlapEMG}E).

\section{Discussion}
\label{discussion}
The 3D reconstructions reveal muscle NMJ distributions that closely resemble those described in the literature \cite{carre2022distribution,cardasis1981ultrastructural,luff1998age}. The estimation of length (18.17 ± 2.69 mm) and volume (87.10 ± 14.99 $mm^3$) of the muscle also show reasonable values within the expectations \cite{mele2016vivo,desplanches1990rat}. However, both these estimation methods have limitations. One key drawback is their inability to compensate for tissue deformation introduced during histological processing, such as folds or tears, which can compromise accuracy. As a result, while this approach provides a reasonable approximation of the true muscle volume and length, it requires validation against measurements obtained through alternative methods. Further, the muscle length estimation relies solely on the 2D histology images, not fully capturing the full muscle’s three-dimensional structure.

The voxel averaging method (Section \ref{voxel}) improves the visualization of spatial distribution patterns and highlights differences across animals within the same model. From our results for example, the naïve model's NMJs exhibit distinct overlap at the muscle center, extending longitudinally with higher density in the upper region, suggesting the presence of a nerve branching and traversing the muscle. This pattern recurs across animals, resulting in a concentrated NMJ region with higher average density. 

Density-based clustering (Section\ref{dbscan}) provides a clearer visualization of the primary regions where features are present, highlighting the morphology and size of clusters more effectively than the voxel averaging method (Section \ref{voxel}). As shown in Figure \ref{fig:results}, the naïve muscle displays a central cluster of recurring NMJs, likely corresponding to the main branch of the tibial nerve, with longitudinal clusters below it that likely represent nerve ramifications. 

The NMJ to hdEMG reconstruction yielded a Pearson correlation coefficient of R = 0.21 (Figure \ref{fig:overlapEMG}E). This is not unexpected given the inherent limitations of the methods employed. First, the Myomatrix electrode is composed of eight individual flexible microelectrode strands, each with eight recording sites. This design renders the electrode placement on the surface variable, and although we approximated the array as a rectangular grid, the actual borders are less well defined, leading to potential misalignment. Additionally, the NMJ counts likely represent only a fraction of the entire muscle, as only one slide out of every eight was used—thus capturing roughly one-eighth of the total tissue—which limits the scope of data extraction. The validation technique is further complicated by tissue atrophy and rupture during the fixing process, both of which can alter the relative dimensions of the recording area compared to the tissue. Despite these limitations, the observed weak correlation indicates that our approach can still provide useful insights. This technique can help facilitate the estimation of correspondences between individual NMJs and their associated motor unit locations. Thus, this method is still effective for studying the structural and electrical properties of the tissue. 

Some limitations of the proposed algorithm include the potential impact of histological artifacts, such as tissue folding or rupture, which can significantly affect the accuracy of 3D reconstructions and measurements, such as volume and dimensions. Additionally, the method does not incorporate automated feature identification, relying instead on manual identification. This approach can be time-consuming and prone to human error, such as mislabeling or missing features on individual slides. In addition, these methods have been tested exclusively in rodent muscle tissue, which requires further validation with other types of muscle tissue of different species. 

Future work should focus on developing refined techniques that better account for the flexible and deformable nature of the electrodes, as well as mitigating the impact of tissue alterations during sample preparation (such as by using block phase imaging \cite{magnain2014blockface}).  If this method were to be used or replicated in the future, we recommend recording epimysial activity using a rectangular array of electrodes on the muscle to maintain cohesiveness with the overlap in the reconstruction. In addition, we recommend flash freezing the tissue, potentially with the electrode if expendable, to maintain better integrity of the tissue and reduce the deformation of the tissue.

The primary goal and strength of this work is to equip researchers performing muscle-related research with a versatile tool for analyzing any structural feature within muscle tissue—even those beyond the NMJs, such as T-tubules, sarcomeres, and vasculature. This approach provides a more complete understanding of feature distribution throughout the tissue and their functional roles when compared with in vivo recordings and electrophysiological data. For instance, this approach can be applied to studies examining tissue regeneration, where the structural arrangement of key features dynamically evolves over time. Additionally, it holds significant promise for investigating muscle-specific disorders—ranging from muscular dystrophies and age-related degeneration to vascular diseases that compromise muscle integrity. By accommodating such diverse applications, this method shows its versatility across a wide spectrum of muscle research fields.

By providing a precise spatial and functional mapping of NMJs, this algorithm can greatly aid studies relying on NMJ analysis to understand neuromuscular adaptations and pathology which is critical for advancing neuromuscular implant technologies that leverage skeletal muscle’s natural bio‐amplification of nerve signals \cite{quinn2024neuromuscular, lowe2024volume}. For example, it can streamline assessments of NMJ remodeling, critical for evaluating interventions aimed at improving functional recovery in models of muscle injury or neurodegeneration. This capability supports the investigation of NMJ integration in regenerative therapies, as explored in studies addressing volumetric muscle loss (\cite{johnson2024combined}), enhances understanding of motor unit remodeling during aging and exercise interventions (\cite{jones2022ageing}), and offers a standardized method to track NMJ degeneration and reinnervation patterns in disease models like ALS (\cite{martineau2018dynamic}). 

In conclusion, this study introduces a novel method for muscle research, focusing on analyzing histology images and their features through 3D reconstruction of the organ, identifying structural patterns, and integrating these with in vivo electrophysiological data to establish a link between structure and function.

\section*{Declarations}
\subsection*{Author contribution statement}
AAO, MB, PLP, SW, KNQ - Conception and design of the study, acquisition of data and analysis and interpretation of data.
\\
KG, FK - Acquisition of data and analysis
\\
NVT - Conception and design of the study and interpretation of data.
\subsection*{Ethical statement}
This study was reviewed and approved by the Johns Hopkins Animal Care and Use Committee.
\subsection*{Data Availability}
Data associated with this study has been deposited at \url{https://github.com/AleAsca/Histology-NMJ-and-muscle-Activity-Toolbox}
\subsection*{Declaration of Interest statement}
The authors declare no conflict of interest.
\subsection*{Declaration of generative AI and AI-assisted technologies in the writing process}
During the preparation of this work the author(s) used ChatGPT 4o in order to improve sentence flow and structural clarity. After using this tool/service, the author(s) reviewed and edited the content as needed and take(s) full responsibility for the content of the publication.
\subsection*{Acknowledgments}
The authors would like to thank the members of the Center for Advanced Motor BioEngineering Research (CAMBER) for providing Myomatrix arrays for this study. We also express our gratitude to Dr. Sami Tuffaha’s lab at Johns Hopkins University for providing us with access to the Leica DMi8 and generously contributing their time and expertise through comprehensive training sessions. Figures created in BioRender. Ascani Orsini, A. (2025) \url{https://BioRender.com/i10r303}
\subsection*{Funding Statement}
This research did not receive any specific grant from funding agencies in the public, commercial, or not-for-profit sectors.
\subsection*{Additional information}
No additional information is available for this paper.

\bibliographystyle{elsarticle-harv} 
\bibliography{main}

\newpage
    \begin{longtable}{@{}p{0.20\textwidth} p{0.65\textwidth} p{0.10\textwidth}@{}}
\midrule
\caption{Summary of major variables used in the analysis of muscle histology, NMJ clustering, and electrode overlap. Variables from \texttt{analysis.m} set the physical scaling and 3D reconstruction parameters, those from \texttt{Avg\_NMJ\_contour.m} pertain to NMJ coordinate extraction, normalization, and clustering, and variables from \texttt{MeshOverlap.m} are used for overlaying electrode data on muscle images. Note that variables sharing the same name (e.g. \texttt{umpx}) may have different numerical values depending on the script context.}
\label{tab:physiological-variables}\\

\toprule
\textbf{Variable} & \textbf{Description} & \textbf{Units} \\
\midrule
\endfirsthead

\multicolumn{3}{c}{\tablename\ \thetable\ -- \textit{Continued from previous page}} \\
\toprule
\textbf{Variable} & \textbf{Description} & \textbf{Units} \\
\midrule
\endhead

\midrule \multicolumn{3}{r}{\textit{Continued on next page}} \\
\endfoot

\bottomrule
\endlastfoot

\multicolumn{3}{l}{\textbf{From \texttt{analysis.m} (Sections \ref{reconstruction} - \ref{length})}} \\[0.5ex]
\midrule
\texttt{slideS}   & Spacing between histological slides (slice thickness)  & mm \\
\texttt{scale}    & Scaling factor for image resizing in 3D reconstruction    & -- \\
\texttt{umpx}     & Conversion factor from \(\mu\)m to pixels (analysis version)   & pixels/\(\mu\)m \\
\texttt{V}        & Estimated muscle volume per slide (nonzero pixels count)  & mm\(^3\) \\
\texttt{Vtot}     & Total muscle volume (sum over slides)                   & mm\(^3\) \\
\texttt{refDist}  & Maximum muscle dimension (longest distance measured)    & mm \\
\texttt{NMJs}     & Coordinates of neuromuscular junctions (x, y, and later z)  & pixels or mm \\
\texttt{type}     & NMJ classification labels (e.g. Types 1--4 or NaN)        & -- \\[1ex]

\multicolumn{3}{l}{\textbf{From \texttt{Avg\_NMJ\_contour.m} (Sections \ref{norm} - \ref{dbscan})}} \\[0.5ex]
\midrule
\texttt{figfile}      & File paths for reconstructed muscle figure files         & File path (string) \\
\texttt{dir\_model}   & Directory for the 3D muscle model file                   & File path (string) \\
\texttt{fname}        & Name of the file containing NMJ coordinates              & String \\
\texttt{scalePoints}  & Scale factor for marker size in NMJ plots                & -- \\
\texttt{umpx}         & Conversion factor from \(\mu\)m to pixels (contour version) & pixels/\(\mu\)m \\
\texttt{size\_vox}    & Voxel size for grid creation (computed as 20 \(\times\) umpx) & pixels \\
\texttt{rnp}          & Radius for neighboring points in DBSCAN clustering       & (coordinate units) \\
\texttt{minp}         & Minimum number of points to form a cluster               & -- \\
\texttt{w}            & Muscle width extracted from figure files                 & pixels or mm \\
\texttt{l}            & Muscle length extracted from figure files                & pixels or mm \\
\texttt{allPoints}    & Cell array of NMJ coordinates from each figure           & (coordinate units) \\
\texttt{refMin}       & Minimum (x,y,z) values across NMJ points (normalization)   & (coordinate units) \\
\texttt{refMax}       & Maximum (x,y,z) values across NMJ points (normalization)   & (coordinate units) \\
\texttt{refLen}       & Reference muscle length (from the first dataset)         & pixels or mm \\
\texttt{refWid}       & Reference muscle width (from the first dataset)          & pixels or mm \\
\texttt{normPoints}   & Normalized NMJ coordinates for each dataset              & -- \\
\texttt{onModel}      & Concatenated, scaled NMJ coordinates (excluding first dataset) & (coordinate units) \\
\texttt{x}, \texttt{y}, \texttt{z} & Grid vectors for voxel analysis              & (coordinate units) \\
\texttt{a}            & 3D voxel count matrix (NMJ occurrence per voxel)         & Count \\
\texttt{avgC}         & Average NMJ coordinate within each voxel (for clustering)  & (coordinate units) \\
\texttt{IDX}          & Cluster indices assigned via DBSCAN to NMJ points        & -- \\
\texttt{denseCluster} & Label of the densest NMJ cluster                         & -- \\
\texttt{selected\_points} & NMJ coordinates belonging to the densest cluster       & (coordinate units) \\[1ex]

\multicolumn{3}{l}{\textbf{From \texttt{MeshOverlap.m} (Section \ref{overlay})}} \\[0.5ex]
\midrule
\texttt{umpx}    & Scaled conversion factor from \(\mu\)m to pixels (mesh overlap version) & pixels/\(\mu\)m \\
\texttt{dT}      & Normalized distance from the top of the muscle           & -- \\
\texttt{sR}      & Size ratio of electrodes relative to the muscle          & -- \\
\texttt{topLR}   & Indicator for top muscle orientation (1: right, 0: left)   & -- \\
\texttt{damp}    & Damping factor for tissue adjustment                      & -- \\
\texttt{showIMG} & Flag to display images during processing                  & Boolean \\
\texttt{hframe}  & Heatmap frame index used for calibration                 & Integer (frame index) \\
\texttt{heatmapExample} & File path for sample heatmap data                   & File path (string) \\
\texttt{muscle}  & Muscle image data extracted from the figure file           & Cell array (image matrices) \\
\texttt{lmax}    & Maximum muscle length (derived from \texttt{l})            & pixels \\
\texttt{LdT}     & Scaled distance from the top (lmax \(\times\) dT)           & pixels \\
\texttt{LsR}     & Scaled length of the electrodes (lmax \(\times\) sR)        & pixels \\
\texttt{cL}      & Center of the muscle length (lmax/2)                       & pixels \\
\texttt{lastR}   & Remaining muscle length after electrode region           & pixels \\
\texttt{distC1}  & Distance from center to top region (cL - LdT)              & pixels \\
\texttt{distC2}  & Distance from center to lower region (cL - lastR)          & pixels \\
\texttt{mask}    & Binary mask defining the electrode region on each slide    & Logical array \\
\texttt{cM}      & Center coordinates of each muscle image (per slide)        & pixels \\
\texttt{rX}, \texttt{rY} & X and Y coordinate ranges for electrode overlay    & Pixel indices \\
\texttt{testH}   & Processed heatmap image (sampled, resized, rotated)        & Matrix (intensity values) \\
\texttt{eVal}    & Electrode value matrix extracted for each slide            & Matrix (intensity values) \\
\texttt{EMGslide} & Electrode overlay on muscle images (eVal \(\times\) mask)   & Matrix \\
\texttt{EMGslidediff} & Difference matrix computed between slides (via centeredSubtraction) & Matrix \\
\texttt{cumEMGdiff} & Cumulative difference matrix for electrode intensity adjustment & Matrix \\

\end{longtable}

\newpage

\end{document}